\documentclass[manuscript]{aastex}
\usepackage{amssymb}

\slugcomment{Not to appear in Nonlearned J., 45.}

\shorttitle{Collapsed Cores in Globular Clusters}
\shortauthors{Djorgovski et al.}

\begin{document}


\title{Long-term monitoring of the TeV emission from Mrk 421 with the ARGO-YBJ experiment}


\author{B.~Bartoli\altaffilmark{1,2},
 P.~Bernardini\altaffilmark{3,4},
 X.J.~Bi\altaffilmark{5},
 C.~Bleve\altaffilmark{3,4},
 I.~Bolognino\altaffilmark{6,7},
 P.~Branchini\altaffilmark{8},
 A.~Budano\altaffilmark{8},
 A.K.~Calabrese Melcarne\altaffilmark{9},
 P.~Camarri\altaffilmark{10,11},
 Z.~Cao\altaffilmark{5}
% ...
\\ (The ARGO-YBJ Collaboration)}


 \altaffiltext{1}{Dipartimento di Fisica dell'Universit\`a di Napoli
                  ``Federico II'', Complesso Universitario di Monte
                  Sant'Angelo, via Cinthia, 80126 Napoli, Italy.}
 \altaffiltext{2}{Istituto Nazionale di Fisica Nucleare, Sezione di
                  Napoli, Complesso Universitario di Monte
                  Sant'Angelo, via Cinthia, 80126 Napoli, Italy.}
 \altaffiltext{3}{Dipartimento di Fisica dell'Universit\`a del Salento,
                  via per Arnesano, 73100 Lecce, Italy.}
 \altaffiltext{4}{Istituto Nazionale di Fisica Nucleare, Sezione di
                  Lecce, via per Arnesano, 73100 Lecce, Italy.}
 \altaffiltext{5}{Key Laboratory of Particle Astrophysics, Institute
                  of High Energy Physics, Chinese Academy of Sciences,
                  P.O. Box 918, 100049 Beijing, China.}
 \altaffiltext{6}{Dipartimento di Fisica Nucleare e Teorica
                  dell'Universit\`a di Pavia, via Bassi 6,
                  27100 Pavia, Italy.}
 \altaffiltext{7}{Istituto Nazionale di Fisica Nucleare, Sezione di Pavia,
                  via Bassi 6, 27100 Pavia, Italy.}
 \altaffiltext{8}{Istituto Nazionale di Fisica Nucleare, Sezione di
                  Roma Tre, via della Vasca Navale 84, 00146 Roma, Italy.}
 \altaffiltext{9}{Istituto Nazionale di Fisica Nucleare-CNAF, Viale
                  Berti-Pichat 6/2, 40127 Bologna, Italy.}
 \altaffiltext{10}{Dipartimento di Fisica dell'Universit\`a di Roma ``Tor  Vergata'',
                   via della Ricerca Scientifica 1, 00133 Roma, Italy.}
 \altaffiltext{11}{Istituto Nazionale di Fisica Nucleare, Sezione di
                   Roma Tor Vergata, via della Ricerca Scientifica 1,
                   00133 Roma, Italy.}
% ...

\begin{abstract}
ARGO-YBJ is an air shower detector array with a fully covered layer of resistive plate chambers. It is operated with a high duty cycle and a large field of view.
It continuously monitors the northern sky at energies above 0.3 TeV. In this paper, we report a long-term monitoring of Mrk 421 over the period from  2007 November to 2010 February. This source was observed by the satellite-borne experiments  $Rossi$ $X$-$ray$ $Timing$ $Explorer$ and $Swift$ in the X-ray band. Mrk 421 was especially active in the first half of  2008. Many flares are observed in both X-ray and $\gamma$-ray bands simultaneously.
The $\gamma$-ray  flux observed by ARGO-YBJ has
a clear correlation with the X-ray flux. No lag  between the X-ray and $\gamma$-ray photons longer than 1 day is found. The evolution of the spectral energy distribution is investigated by measuring spectral indices at four different flux levels. Hardening of the spectra is observed in both X-ray and $\gamma$-ray bands. The $\gamma$-ray flux increases quadratically with the simultaneously measured X-ray flux.
All these observational results strongly favor the synchrotron self-Compton process as the underlying radiative mechanism.
\end{abstract}


\section{Introduction}
Mrk 421 ($z=0.031$) is one of the brightest blazars known and is classified as a BL Lac object, a subclass of active galactic nuclei (AGNs).
 Mrk 421 was the
first BL Lac source detected (by EGRET in 1991) at energies
above 100 MeV \citep{lin92}, and was also the first extragalactic object detected by a ground-based
experiment (Whipple) at energies around 1 TeV \citep{punch92}(in the following we will refer to $\gamma$-rays as  those around 1 TeV). Its emission, like that of the
other blazars, is generally dominated by nonthermal
radiation from a relativistic jet   aligned along our line of
sight. The spectral energy distribution (SED)  is double-humped at X-ray  and  $\gamma$-ray energies in a plot of $\nu F_{\nu}$ versus
$\nu$ \citep{fossati98}, where $\nu$ is the frequency and $F_{\nu}$ the flux density. The  hump at low energies is usually
interpreted as being due to synchrotron radiation from relativistic
electrons (and positrons) within the jet. The origin  of the
hump at high energies is under debate. Many models attribute the high-energy emission to the
inverse Compton scattering of the synchrotron (synchrotron
self-Compton, SSC) or external photons (external Compton, EC) by the
same population of relativistic electrons \citep{ghise98, dermer92},
 therefore an X-ray/$\gamma$-ray correlation would naturally be expected. Other
models invoke hadronic processes including
proton-initiated cascades and/or proton-synchrotron emission in a
magnetic-field-dominated jet. Although the hadronic models may also
accommodate the observed SED and X-ray/$\gamma$-ray correlation
\citep{aharo00, mucke03}, they are generally challenged by the most
rapid flares in the TeV region \citep{gaidos96}.

 Mrk 421 is a
very active blazar with    major outbursts about once every two years in both X-ray  \citep{cui04} and $\gamma$-ray  \citep{tlucz10} bands. A major outburst usually lasts
several months and is accompanied by many rapid flares with
timescales from tens of minutes to several days. Its high variability and
broadband emission require  long-term, well-sampled, multiwavelength
observations in order to understand the emission mechanisms of these outbursts.
During the last decade, several coordinated multiwavelength campaigns focusing on Mrk 421 have been conducted both in response to strong outbursts and as part of dedicated observation campaigns \citep{rebillot06, fossati08,
acciari09, donna09,horan09}. Some important general features of the AGN flares have been obtained. Although X-rays and $\gamma$-rays are found  to be strongly correlated,
neither type is  evidently correlated with optical and radio emissions. The spectral index becomes harder at higher fluxes in both X-ray and $\gamma$-ray
bands \citep{rebillot06,krenn02,aielli10}. An intensive
multiwavelength monitoring campaign has  recently been conducted with the
Whipple telescope and the $Rossi$ $X$-$Ray$ $Timing$ $Explorer$ ($RXTE$) \citep{blaze05}.
Similar features, including correlated variability at different energies, flaring and spectral evolution are also observed.
All these phenomena can be interpreted in the
framework of the SSC model. However, ``orphan flares'', which have only $\gamma$-ray emission without low-energy companions, and a lag of about two days between X-rays and $\gamma$-rays \citep{blaze05} are usually recognized as major challenges to the model.

A long-term simultaneous  X-ray/$\gamma$-ray
observation  is better performed   by means of a combination of
 satellite-borne X-ray experiments and wide field-of-view air shower
experiments, such as the Tibet AS-$\gamma$ experiment \citep{ameno03} and ARGO-YBJ experiment \citep{aielli06}, which are operated  day and night with a duty cycle higher than 85\% and can observe any source with a zenith angle less than 50$^{\circ}$. This is essential in order to investigate the temporal features of AGN emissions.  The ARGO-YBJ experiment has continuously monitored the northern sky for outbursts from all AGNs, such as Mrk 421, since  2006 June. Meanwhile, these sources were also monitored by the satellite-borne X-ray detectors All-Sky Monitor (ASM)/$RXTE$ and  Burst Alert
Telescope (BAT)/$Swift$.
In this paper, we report on the long-term monitoring of Mrk 421 for $\gamma$-ray outbursts and on the correlation between  $\gamma$-rays and simultaneous X-rays over the period from   2007 November to   2010 February.
The paper is organized as follows: the ARGO-YBJ experiment is
briefly introduced in Section 2 and its long-term performance is shown
in Section 3. A data analysis method is
described in Section 4.  Observation findings  are presented in Section 5.
 Conclusions are given in Section 6.

\section{The ARGO-YBJ Experiment}
The ARGO-YBJ  experiment, located in Tibet, China at an altitude of 4300 m
a.s.l., is the result of a collaboration among  Chinese and Italian
institutions and is designed for very high energy $\gamma$-ray astronomy and cosmic ray
observations. The detector consists of a single layer of resistive plate chambers (RPCs), which are organized with a modular configuration.
The basic module is a
cluster (5.7 m $\times$ 7.6 m)  composed of 12 RPCs
(2.850 m $\times$ 1.225 m  each). The RPCs are equipped with pick-up
strips (6.75 cm $\times$ 61.80 cm  each), and the logical OR of the signal from eight neighboring strips
constitutes a logical pixel (called a  ``pad'') for triggering and timing
purposes. One hundred thirty clusters are installed to form a carpet
of about 5600 m$^{2}$ with an active area of $\sim$93\%. This central
carpet is surrounded by 23 additional clusters (a ``guard ring'') to
improve the reconstruction of the shower core location. The total area of the
array is  110 m $\times$ 100 m. More details about the detector and
 RPC performance can be found in, for example, \citet[][]{aielli06}.

The RPC carpet is connected to two independent data acquisition
systems  corresponding to two different operation modes, referred to as the shower and the scaler \citep{aielli08}
 modes. Data used in this paper  refer  to the
shower mode, in which the ARGO-YBJ detector is triggered when at least 20 pads in the entire carpet
detector are registered within 420 ns.
 The high granularity of the apparatus permits a detailed
spatial$-$temporal reconstruction of the shower profile and therefore  the incident direction of the primary particle.
The arrival time of the particles is measured by time to
digital converters (TDCs) with a resolution of approximately 1.8 ns.
 In order to calibrate the 18,360 TDC channels, an
off-line method \citep{he07} has been developed using cosmic ray
showers. The calibration precision is 0.4 ns, and the procedure is applied every month
\citep{aielli09}.

The central 130 clusters began taking data in  2006 June, and the
``guard ring'' was merged into the DAQ stream in  2007 November. The
trigger rate is $\sim$3.6 kHz with a dead time of 4\%, and the
average duty cycle is higher than $85\%$.

\section{Detector Performance}
For long-term monitoring campaigns, the stable operation of the
equipment is very important. In order to continuously monitor  the performance of the RPCs,
including detection efficiency and time resolution, a cosmic ray muon telescope is set up near the
detector array.
The RPC efficiency fluctuates by about 0.3\% and the time resolution by
 about 0.4 ns in a day, and these values become 1.5\% and 1 ns in a year, respectively. Detailed information about the performance monitored using this telescope can be found in \citet[][]{aielli09b}.

To estimate the angular resolution and effective area, a full Monte Carlo simulation of the RPC detector array is developed. In the code, the CORSIKA package \citep{heck98} is used to describe the air shower development.
  G4argo \citep{guo10}, a GEANT4-based \citep{geant} package, is used to simulate the
response of the RPC array.  For events with a number of fired pads ($N_{pad}$) greater than 100, the Point Spread Function (PSF) has a single Gaussian functional form. For events at lower N$_{pad}$, the best fit to the PSF becomes a combination of two Gaussian distributions, the wider of which contains 20\% of the events. To simplify the description of the PSF, a parameter $\psi_{70}$ is defined as is the opening angle  containing 71.5\% of the events.
When the PSF is a single Gaussian,  $\psi_{70}$  maximizes the signal-to-background ratio for a point source. For $N_{pad}>1000$, $\psi_{70}$ is
0.47$^{\circ}$, while at $N_{pad}\sim 20$ $\psi_{70}$ becomes 2.8$^{\circ}$.
 The  effective area of the detector for $\gamma$-induced showers depends on the $\gamma$-ray energy and  incident zenith angle, e.g., it is about 100 m$^{2}$ at 100 GeV and $>$10,000 m$^{2}$ above 1 TeV for a zenith angle of 20$^{\circ}$ \citep{aielli09a}.

The angular resolution, pointing accuracy and stability of the ARGO-YBJ detector array have
 been thoroughly tested by measuring the shadow of the Moon in cosmic rays \citep{iuppa09}.
 The shadow is detected with a significance of 10 $\sigma$ per month  using the ARGO-YBJ data. The position of the shadow allows the investigation of
 any pointing bias. The east-west displacement is in good agreement with the expectation,
  while a 0.2$^{\circ}$ pointing error toward  the north is observed and is under investigation.

\section{Data Analysis}
For the analysis presented in this paper, only events with a zenith angle less than 45$^{\circ}$ are used, and the data set  is divided into six groups according to $N_{pad}$. The  event selections  are listed in Table 1,
  where $R$ is the distance between shower core position and the carpet center, and TS
  is the time spread of the shower front in the conical fit defined in Eqation(1) of \citet[][]{aielli09}. With these selections,
  the angular resolution is improved, e.g., for events with $N_{pad}>60$ and $N_{pad}>100$, the  opening angle $\psi_{70}$ decreases from 1.68$^{\circ}$ and 1.27$^{\circ}$ to 1.36$^{\circ}$ and 0.99$^{\circ}$. As a consequence, the significance of the Crab Nebula is increased by about 10\%  and 25\%, respectively.

 In order to obtain a sky map using events in each $N_{pad}$ group, an area centered at the source location in celestial coordinates (right ascension and declination) is divided into a grid of $0.1^{\circ}\times0.1^{\circ}$ bins and filled with detected events according to their reconstructed origin. The number of events in each grid bin is denoted as $n_{i}$, where the subscript $i$ denotes the bin number. In order to extract an excess of $\gamma$-rays from the source, the  direct integral method 
\citep{fleysher04} is applied to estimate the number of cosmic ray background events in the bin, denoted as $b_{i}$. An essential assumption in this estimation is that the background must be uniform around the source. However, an anisotropy of the cosmic ray flux is measured over spatial scales such as $10^{\circ}\times10^{\circ}$ and larger \citep{ameno06,zhangjl09}. This anisotropy as measured by the $n_{i}/b_{i}$ ratio is stable; therefore, it is possible to correct it with a long-term measurement for each grid bin. An average of the ratio over  the bins in a window $11^{\circ}\times11^{\circ}$  centered on the source bin is applied for smoothing. In this procedure, in order to avoid any contamination of the excess in the source bin and possible spread out due to the finite angular resolution, the contribution from a $5^{\circ}\times5^{\circ}$ window around the source bin is excluded. Finally the correction factor, denoted as $\beta_{i}$, is calculated as follows:
\begin{equation}
\beta_{i}=\frac{1}{m}\sum\limits_{j=1}^{m}  \frac{n_j}{b_j} ,
\end{equation}
where the subscript $j$ is the index of the $m=12100-2500=9600$ selected grid bins. The corrected number of background events in each bin is $b_{i}^*=\beta_{i} b_{i}$. The typical value of $\beta$ around Mrk 421 is approximately 0.9995. The value of $\beta$ for each bin is calculated using about two years of data and is stored in a database for routine analysis.

Taking into account the PSF of the ARGO-YBJ detector, the events in a circular area centered on the bin with an angular radius of $\psi_{70}$ are summed together. Namely,
\begin{equation}
N_{on}=\sum\limits_{i=1}^{k} n_{i}, ~~~~~~~N_{b}=\sum\limits_{i=1}^{k} b^*_{i},
 \end{equation}
where $k$ is the number of bins in the circular area,  $N_{on}$ is the total number of events, and $N_{b}$ is the number of background events. The Li$-$Ma formula \citep{li83} is used to estimate the significance.

\section{Results}
\begin{figure}
\epsscale{.80} \plotone{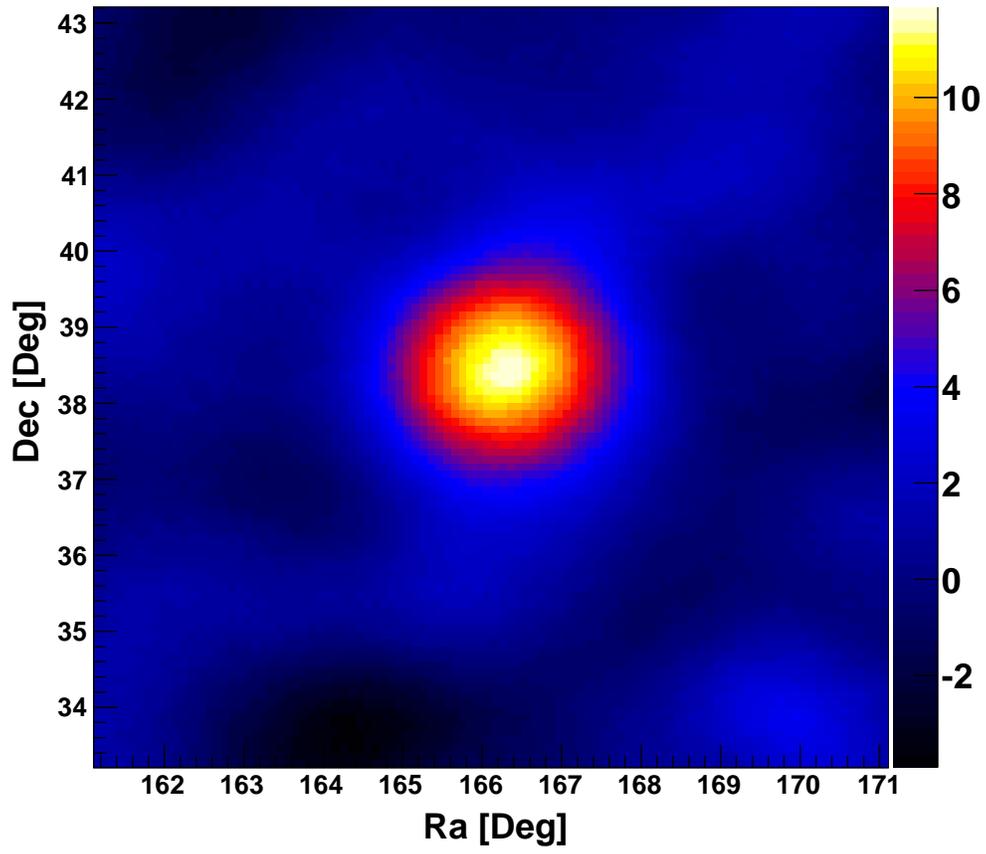}
 \caption{ Distribution of statistical significance around Mrk 421.\label{fig2}}
\end{figure}

The data used in this paper were collected by the ARGO-YBJ experiment
in the period from 2007 November to 2010 February. The total
lifetime is 676.0 days. The numbers of events in different groups after  the selections  are listed in Table 1. A clear signal from Mrk 421 with significance greater than 11$\sigma$ is observed using events with  $N_{pad}>60$ (see
Figure~\ref{fig2}).
A signal at such a level of significance allows us to study flux variations, correlations with the X-ray flux, and the evolution of the SED.

\subsection{Temporal Analysis}
In order to study the correlation between $\gamma$-rays and X-rays, the daily averaged light
curves of both the hard X-rays (15$-$50 keV) measured by BAT/$Swift$\footnote{ Transient monitor results provided by the BAT/$Swift$ team:
\url{http://heasarc.gsfc.nasa.gov/docs/swift/results/transients/weak/Mrk421/}.}
 and the soft X-rays (2$-$12 keV) measured by ASM/$RXTE$\footnote{Quick-look results provided by the ASM/$RXTE$ team:
\url{http://xte.mit.edu/ASM\_lc.html}.} are used.
The observations by $RXTE$ and $Swift$ have a rather long exposure by orbiting the Earth every 1.5 hr.
Since the fluctuation of the X-ray
flux is abnormally large in some days,
in order to control the quality of the data, days that have a very large error on the mean daily event rate are removed from the data set.
For ASM/$RXTE$, the distribution of the error indicates that a selection of the errors smaller than 1 count s$^{-1}$ will cut everything
beyond four standard deviations in the distribution. A similar cut applies to the BAT/$Swift$ data, in which  a selection of the errors smaller
 than 0.0035 counts cm$^{-2}$ s$^{-1}$   cuts everything beyond four standard deviations in the error distribution.
  Approximately, 6.4\% and 5.6\% of events are removed from the $RXTE$ and $Swift$ data sets, respectively.
  Whether it is day   or night, ARGO-YBJ observes Mrk-421 while the AGN is in its field of view.
  A typical transit lasts usually 6 hr. An observational time less than 5 hr  day$^{-1}$ indicates some malfunctioning of the detector in that day,
   which is thus removed from the data set. In total, 9.7\% of data are removed in this way. Finally, 737, 728, and 712 days are selected from the ASM, BAT, and ARGO-YBJ reconstructed data sets, respectively.

\subsubsection{Light Curves}
In 552 days all  three experiments observed Mrk 421 simultaneously. In Figure~\ref{fig3}, the accumulation of event rates from the Mrk
421 direction is shown. The $Swift$ event rate has been normalized using the $RXTE$ scale
and the ARGO-YBJ curve is obtained using events with $N_{pad}>100$, thus the median energy of the observed photons is  1.8 TeV, assuming a spectral index
$-$2.4.
The fast increase in the three curves indicates that the source had a long-term outburst at the beginning of 2008. The following quiet state lasted for about 200 days. Afterward  Mrk 421 became increasingly  more active.  In fact, there were flares
 in 2009 November \citep{isobe10}. The duty cycle of ARGO-YBJ was low due to detector maintenance, therefore it
is not obvious in Figure~\ref{fig3}. There was a large flare in  2010 February \citep{isobe10,ong10}.

Out of the long-term variation that is clearly revealed in the cumulative light
curve shown in Figure~\ref{fig3}, Mrk 421  undergoes a large outburst
during the period from 2008 February to June, indicated by the steepest part of the curves. In fact, it is a combination of
several large flares. A better view of these is shown in Figure~\ref{fig4}, where a smoothing analysis is applied for both $\gamma$-ray and X-ray curves, and each point is the event rate averaged over
five days.   Four large
flares are observed by all   three detectors, and the peak times  are in
good agreement with one another. The fourth flare has been reported by the ARGO-YBJ experiment in \citet[][]{aielli10}.
It gives an important observation when the Cherenkov telescopes are hampered by the Moon.
It can be concluded  that there exists a good long-term correlation
between $\gamma$-rays and X-rays.

\subsubsection{X-ray/TeV Correlation}
The discrete correlation function (DCF)
\citep{edelson88} is used to quantify the degree of correlation and the
phase differences (lags) in the variations between $\gamma$-rays and X-rays.
The daily fluxes before smoothing are used for this analysis.
The DCF (in 1 day
bins) derived from $RXTE$ and ARGO-YBJ data (with $N_{pad}>100$) is shown in the
left panel of Figure~\ref{fig5}, where a positive value means that
$\gamma$-rays lag X-rays.  The peak of the distribution is around zero and the
correlation coefficient at zero is $\simeq0.77$. The result
derived from $Swift$ and ARGO-YBJ data is shown in the right panel of
Figure~\ref{fig5} and the correlation coefficient at zero is $\simeq0.78$. To estimate the lag and its uncertainty, a data-based simulation  suggested by \citet[][]{peter98} is applied and the correlation coefficient between $-$10 and 10 days is fitted with a Gaussian function.
The median value and corresponding 68\% confidence level errors are $-0.14_{-0.85}^{+0.86}$ and $-0.94_{-1.07}^{+1.05}$ days for the correlations of ARGO-YBJ/$RXTE$ and ARGO-YBJ/$Swift$ data, respectively.
 No significant lag longer than one day is found.

\subsection{Spectral Energy Distribution}
To study the SED at different flux levels, the data simultaneously observed in $\gamma$-ray and X-ray bands are divided into four groups according to the observational time periods in which the  ASM/$RXTE$ counting rate is
  $0-2$, $2-3$, $3-5$ or $>5$ cm$^{-2}$ s$^{-1}$. For each
group, a flux-averaged SED is constructed both at $\gamma$-ray and
X-ray energies.

\subsubsection{X-ray Spectra}
ASM/$RXTE$ monitors  the X-ray emission from Mrk 421 at three energy bands, i.e.,
$1.5-3$, $3-5$ and $5-12$ keV \citep{levine96}.
In the flux estimation, the hydrogen column density $1.38\times10^{20}$ cm$^{-2} $\citep{dickey90} and a power law spectrum
are assumed. The best-fit spectral indices for
the four flux levels are $-2.43\pm0.04$, $-2.15\pm0.03$,
$-2.05\pm0.03$, and $-2.02\pm 0.08$, respectively, in which only
statistical errors are taken into account. The spectral indices versus the
corresponding fluences at 10 keV are shown in Figure~\ref{fig6}.  This result is consistent with the analysis of  \citet[][]{rebillot06}, in which  a spectral
hardening toward  high fluxes is also reported based on a shorter timescale observation. This indicates that this correlation is independent of the timescale.

\subsubsection{$\gamma$-ray Spectra}
To estimate the spectrum of $\gamma$-rays with a distribution of the number of events in excess as a function of $N_{pad}$, we follow a widely used method that is described in detail elsewhere \citep{ameno09,aielli10}.  In this procedure, we assume for the spectrum of Mrk 421 a power law with a cutoff factor $e^{-\tau(E)}$, which takes into account the absorption of $\gamma$-rays in the extragalactic background light. We adopt  the optical depth $\tau$(E)  estimated by \citet[][]{franc08}. The ARGO-YBJ detector response is also taken into account. The simulated events are sampled in the energy range from 10 GeV to 100 TeV.

To test this method, the same analysis is performed with the data in the direction of the Crab Nebula, the standard candle in the $\gamma$-ray sky.
The resulting spectrum is
$(4.2\pm0.4_{stat})\times10^{-11}$(E/TeV)$^{-2.57\pm0.09_{stat}}$
photons TeV$^{-1}$ cm$^{-2}$ s$^{-1}$, which is in  agreement with our previous measurement \citep{aielli10} and  observations by
other detectors, such as H.E.S.S. \citep{aharo06}, MAGIC \citep{albert08}, and Tibet AS-$\gamma$ \citep{ameno09}.

Applying this procedure to Mrk 421, we obtain the spectra for
the four event groups with different flux levels. The spectral indices in the energy range from 300 GeV to 10 TeV are $-2.48\pm0.22$, $-2.53\pm0.21$,
$-2.15\pm0.18$,  and $-1.87\pm 0.21$, respectively. Only
 statistical error is quoted.
 The corresponding flux above 1
TeV ranges from 0.8 to 6 times that of the Crab Nebula unit, i.e., $2.67\times10^{-11}$ photons cm$^{-2}$ s$^{-1}$. The spectra seem to become harder with increasing flux, as indicated in Figure~\ref{fig7}, in agreement with the function obtained by the
Whipple experiment \citep{krenn02}. A similar result has been reported elsewhere \citep{aielli10} using the three-day flare data in  2008 June. The quoted errors in Figure~\ref{fig7} are statistical. The systematic error is estimated to be $\lesssim$30\% in the flux level determination \citep{aielli10}.

\subsubsection{Correlation Between $\gamma$-ray and X-ray Fluxes}
 Using the spectra described above, we
investigate the  correlation  between  $\gamma$-ray and X-ray fluxes. Figure~\ref{fig8}
shows the integral $\gamma$-ray flux above 1 TeV as a function of the
 integral X-ray flux from 2 keV to 12 keV; a positive correlation is observed.
A quadratic fit (with the function $y = ax^{2} + b$ ) to the data points yields
$\chi^{2}/$dof$=1.9/2$, while a linear fit yields
$\chi^{2}/$dof$=7.7/2$, where dof refers to degrees of freedom.
The observation  favors a
quadratic  correlation between  $\gamma$-ray and X-ray fluxes.
A similar quadratic correlation has been reported by \citet[][]{fossati08}. In contrast, an observation with linear correlation is
  obtained by \citet[][]{ameno03}.
 According to \citet[][]{katar05}, changes of the magnetic field,
  electron density, and adiabatic cooling may be associated with different correlations between $\gamma$-ray and X-ray fluxes.

\subsection{Modeling of the X-ray and $\gamma$-ray Emissions}
A fit to the four flux-averaged SEDs with a homogeneous
one-zone SSC model proposed by \citet[][]{MaK95} \citep[see
also][]{MaK97,yanget08} is performed. In this model  the parameters include the
Doppler factor $\delta=1/[\Gamma(1-\beta\cos\theta)]$; the spherical
blob radius R; magnetic field strength $B$; electron spectral
index $s$; electron maximum Lorentz factor $\gamma_{\rm max}$; and
electron injection compactness $l_e=\frac{1}{3}m_ec\sigma_{\rm
T}R^2\int_1^\infty d\gamma(\gamma-1)Q_e$, where $\Gamma$ and
$c\beta$ are the Lorentz factor and the speed of the blob,
respectively, $\gamma$ is the electron Lorentz factor,
$\sigma_{\rm T}$ is the Thomson cross section,
 $\theta$ is the angle between its direction of motion
and the line of sight of the observer, and $Q_e$, the  electron spectrum at injection, is assumed to follow a   power law
$Q_e=q_e\gamma^{-s}\exp(-\gamma/\gamma_{\rm max})$. The best fits are shown
in Figure~\ref{fig9} for different flux levels, with
the corresponding parameters given in Table 2. In our fits, the magnetic field strength is estimated by  $B=5\times 10^{-3}\delta \nu_{\rm s,18}\nu^{-2}_{\rm c,27}$ (see Equation (6) of \citet[][]{MaK97}), where $\nu_{\rm s,18}$ is the synchrotron peak frequency in units of $10^{18}$ Hz and $ \nu_{\rm c,27}$ is the IC peak frequency in units of $10^{27}$ Hz. For the lowest flux level (see Figure~\ref{fig9}(1)), the magnetic field strength is estimated to be $\sim 0.08$ G. Compared to the lowest flux level, X-ray peak frequencies in other flux levels increase by a factor of $\sim 2$, but the IC peak frequencies have few changes. Therefore, the magnetic field strengths in other flux levels are larger than that in the lowest flux level by a factor of $\sim 2$ when the Doppler factor $\delta$ is roughly fixed.

\section{Discussion And Summary}
\begin{figure}
\epsscale{.80} \plotone{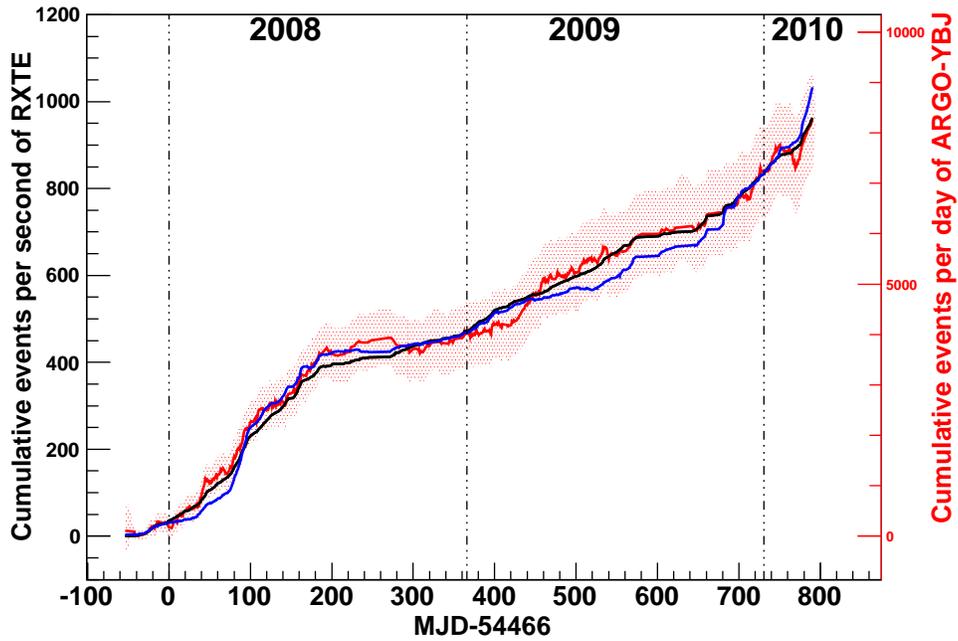} \caption{Cumulative
light curves from the Mrk 421 direction. The red curve is the $\gamma$-ray result observed by ARGO-YBJ, and the shaded red region indicates the corresponding 1$\sigma$ statistical error; the black curve represents soft X-rays (2$-$12 keV) observed by ASM/$RXTE$. Hard X-rays (15$-$50 keV) observed by BAT/$Swift$ are given by the blue curve, and the scale has been normalized to the ASM/$RXTE$ one.\label{fig3}}
\end{figure}

\begin{figure}
\epsscale{.80} \plotone{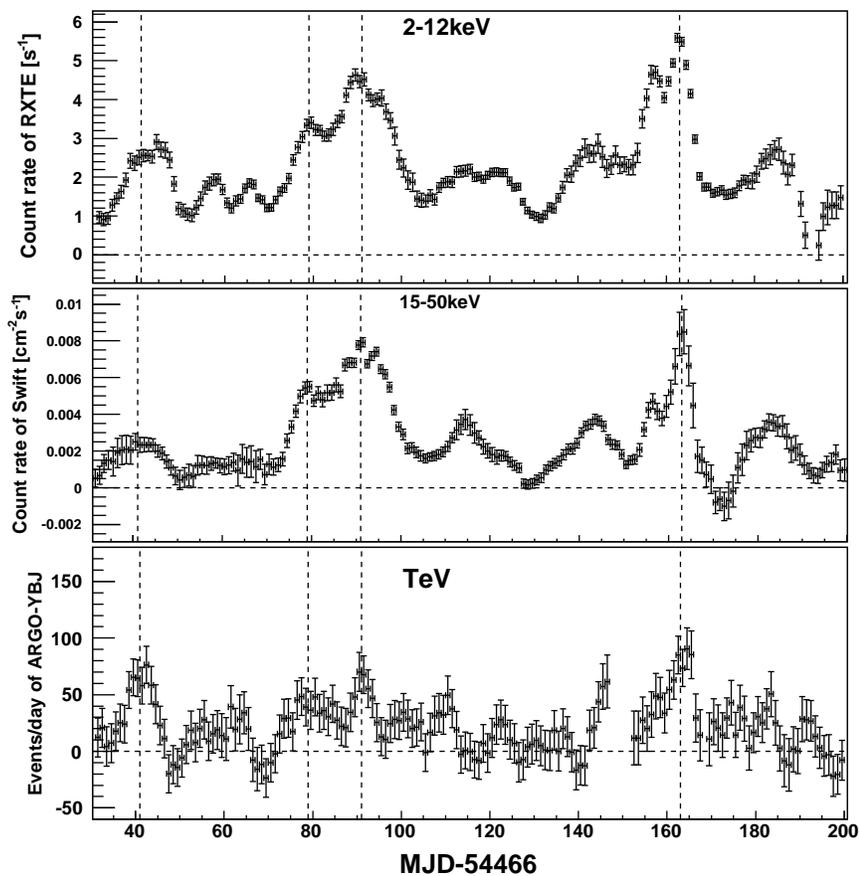} \caption{Daily light
curves from Mrk 421 direction in different energy bands from  2008 February 1 to July 18. Each bin contains the event rate averaged over the five-day interval centered on that bin. The panels from top to bottom refer to 2$-$12 keV (ASM/$RXTE$),
 15$-$50 keV (BAT/$Swift$), and $\gamma$-ray (ARGO-YBJ),
respectively.\label{fig4}}
\end{figure}

Mrk 421 is a very active blazar with frequent
 outbursts, which are composed of many flares and can last as long as a few months.
This makes this blazar an excellent
candidate for studying the jet physics in AGNs. A strong correlation
between its $\gamma$-ray and X-ray emissions has been confirmed by many observations in the past decade (for a review see \citet[][]{wagner08}).
 Most of the previous $\gamma$-ray observations, however, are carried out by
Cherenkov telescopes with limited exposure and
usually focus  on short timescales.
In contrast, the high duty cycle of the ARGO-YBJ experiment makes possible a long-term and continuous observation
of  this variable source, allowing   simultaneous monitoring of  $\gamma$-rays and X-rays for about two years.
 This increases the set of long-term simultaneous multiwavelength
observations of Mrk 421, which are essential for studying the correlation
between energy bands where different emission mechanisms are at work.
The observation time, from  2007 December
to 2010 February, covers both active and quiet phases. The $\gamma$-ray
flux shows a good long-term
correlation with the X-ray flux (see Figure~\ref{fig3}) and all the large X-ray
flares have their $\gamma$-ray counterparts during the outburst
time (see Figure~\ref{fig4}), indicating that $\gamma$-rays and X-rays
may have a common origin as assumed in the SSC model \citep{MaK95}.

\begin{figure}
\epsscale{1.10}
\plottwo{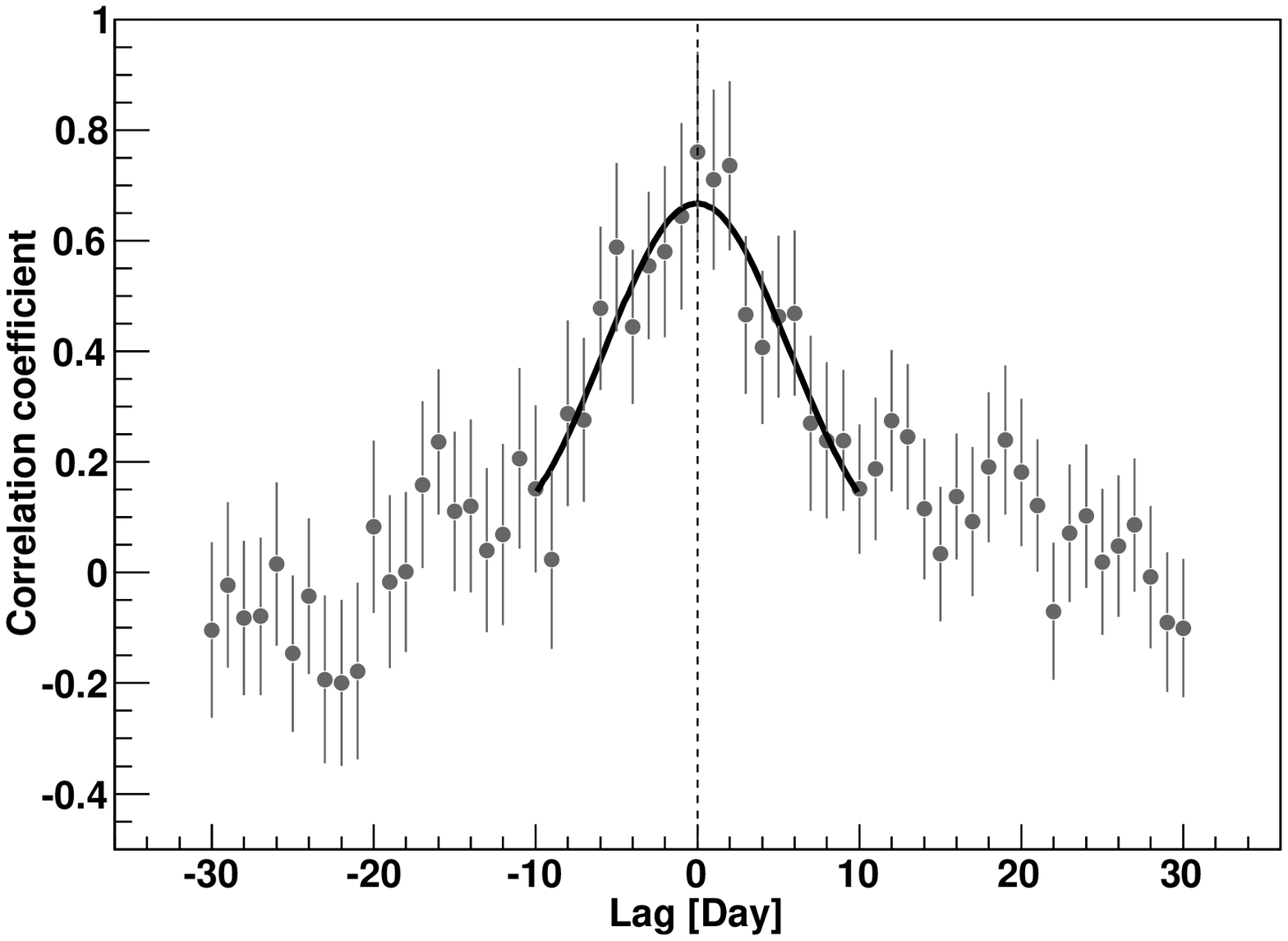}{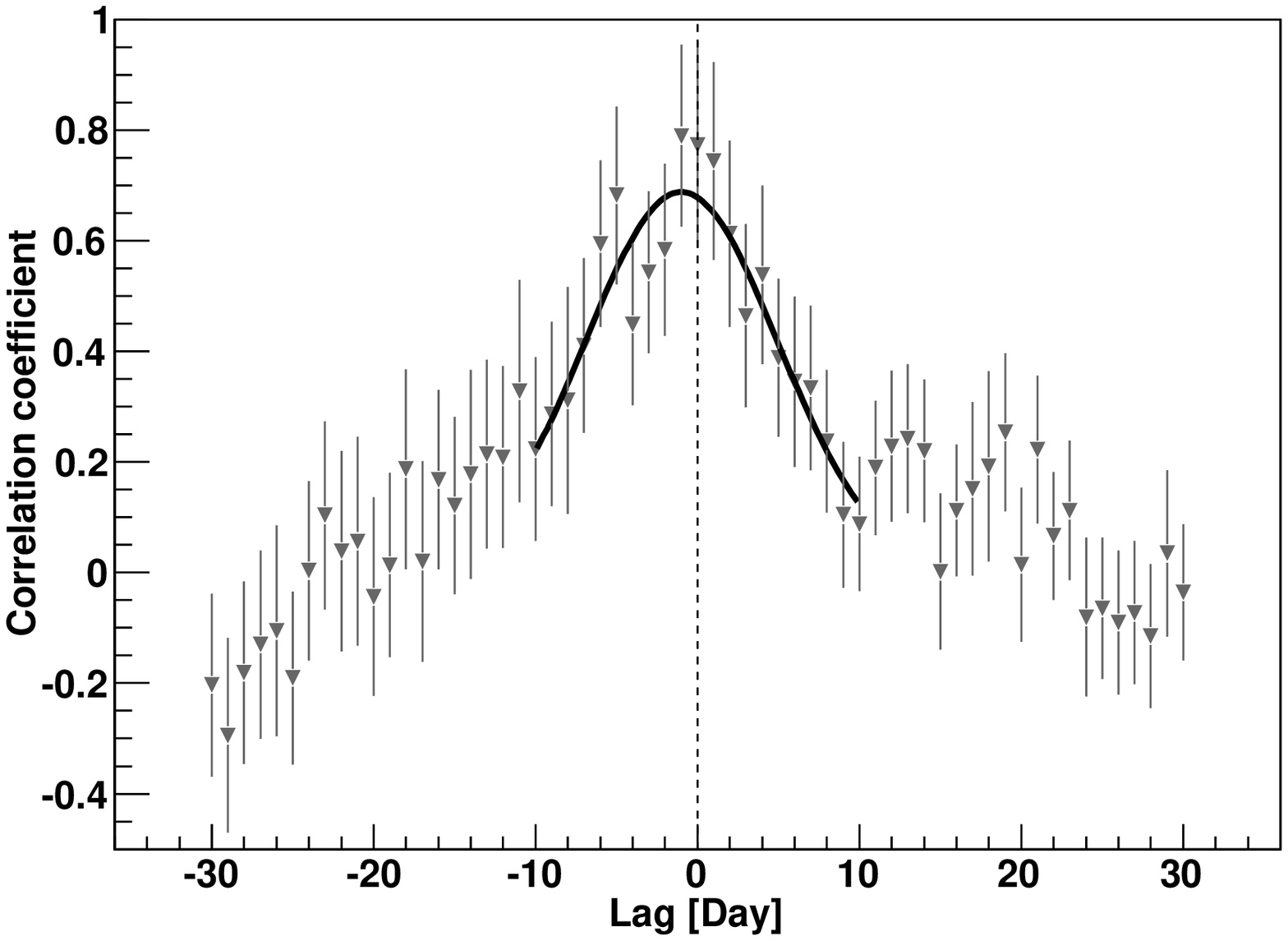}
 \caption{Discrete correlation function between X-ray and $\gamma$-ray light curves  from  2007 November to 2010 February.
  Left: 2$-$12 keV (ASM/$RXTE$) vs. $\gamma$-ray (ARGO-YBJ); a Gaussian function is used to fit from  $-$10 to 10 days.
 Right:  15$-$50 keV (BAT/$Swift$) vs. $\gamma$-ray (ARGO-YBJ).
 Positive value means that $\gamma$-rays lag X-rays. \label{fig5}}
\end{figure}

In the SSC model, the $\gamma$-ray photons are produced via inverse Compton scattering off the synchrotron photons by the same electrons,
and simultaneous variability or short lags are expected between $\gamma$-ray and X-ray fluxes. Short lags can be caused by differences in acceleration and cooling timescales or by reverse shocks, and sub-hour lags have been definitely measured between different X-ray \citep{ravas04} and $\gamma$-ray \citep{albert07} interband energies and between X-ray and $\gamma$-ray bands \citep{fossati08}. On the other hand, the characteristic timescale of the SSC process would be too short to account for a  lag of two days such as that reported by \citet[][]{blaze05} with a marginal significance.  In this paper, the two-year data are used to search for possible lags between variations in X-rays and $\gamma$-rays. No lag longer than one day is observed (see Figure~\ref{fig5}).

\begin{figure}
\epsscale{.80} \plotone{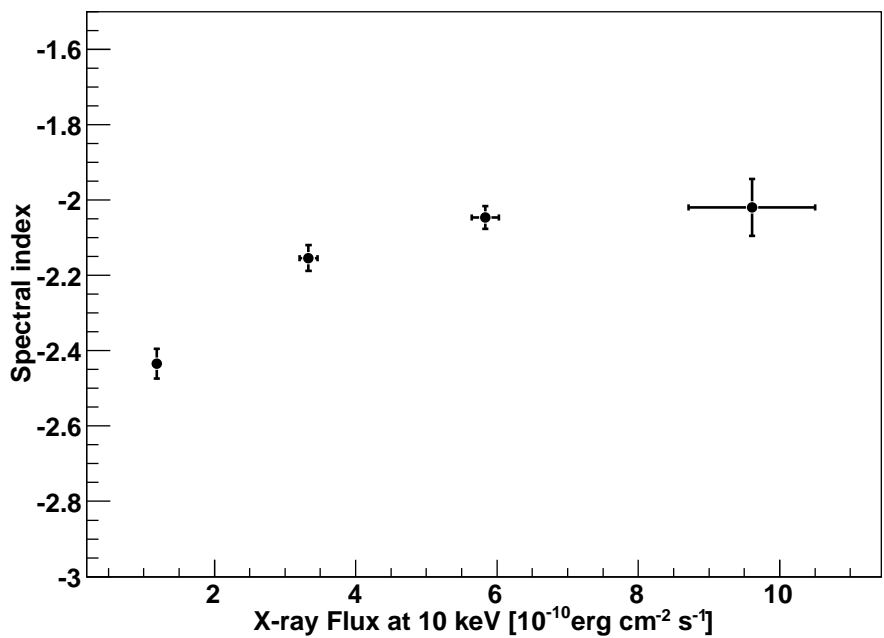} \caption{Correlation between the X-ray
flux at 10 keV and the corresponding photon index at 2$-$12 keV.\label{fig6}}
\end{figure}

\begin{figure} \epsscale{.80} \plotone{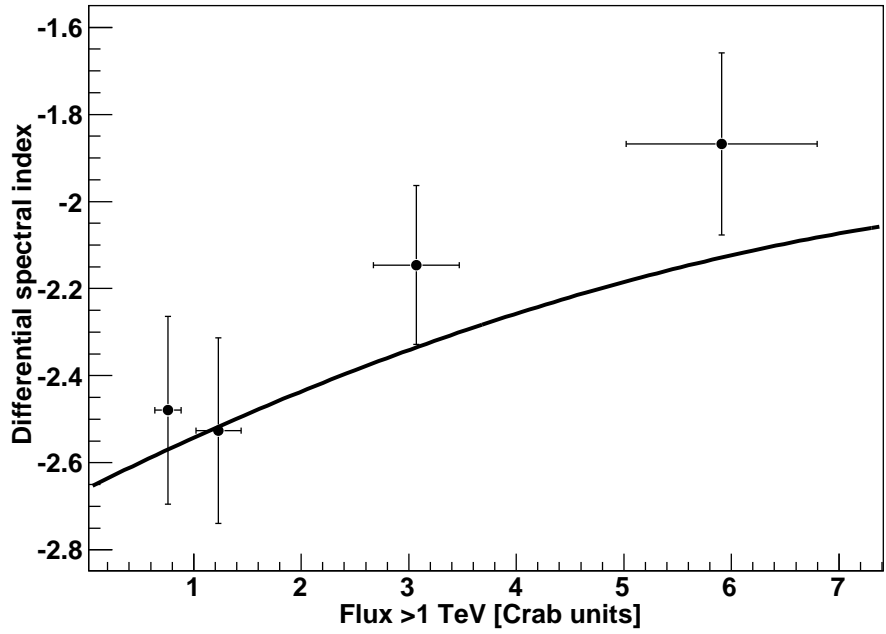} \caption{Spectral index
vs. $\gamma$-ray flux  above 1 TeV. The solid line is the function obtained
by the Whipple experiment \citep{krenn02}.\label{fig7}}
\end{figure}

\begin{figure} \epsscale{.80} \plotone{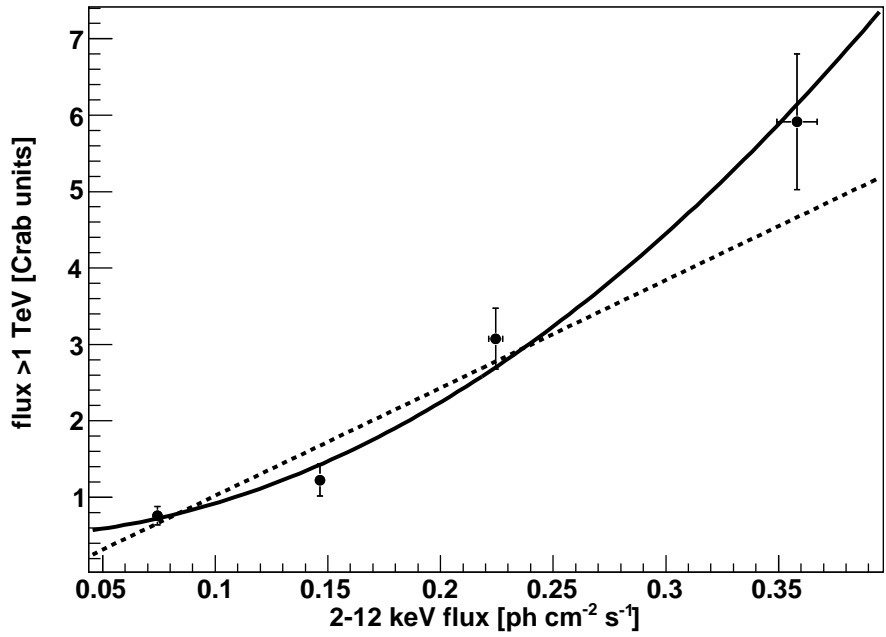} \caption{ $\gamma$-ray  flux above 1 TeV vs. X-ray flux at 2$-$12 keV. The solid line is a quadratic fit using function $y=ax^{2}+b$,
which yields
$\chi^{2}/$dof$=1.9/2$. The dotted line is a linear fit, which
yields $\chi^{2}/$dof$=7.7/2$, where dof refers to degrees of
freedom.\label{fig8}}
\end{figure}

\begin{figure} \epsscale{1.10} \plotone{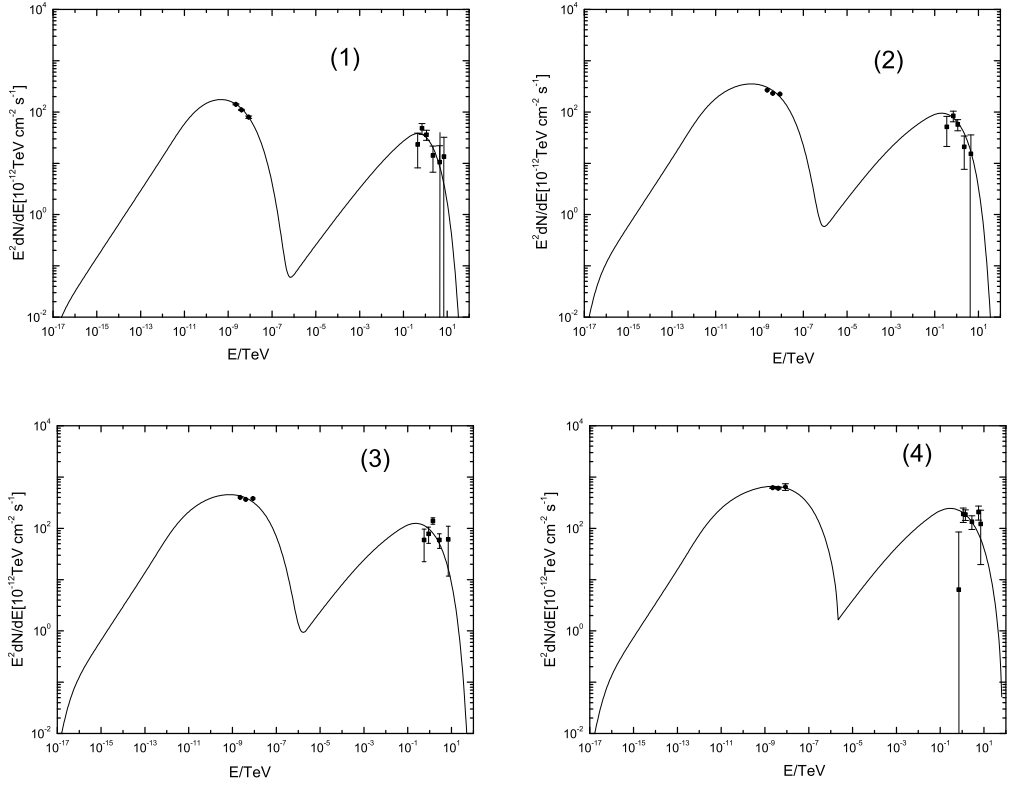}
 \caption{Spectral energy distribution of Mrk 421.
(1) to (4) are derived from four flux level data groups from low to high  according to the ASM/RXTE counting rate (see the text for details).
The solid line
shows the best fit to the data with a  homogeneous one-zone SSC model, and the
best-fit parameters are listed in Table 2. \label{fig9}}
\end{figure}

A sudden variation of flux can be caused by different reasons, e.g., a change in the number of emitting electrons and/or the maximum momentum of emitting electron and/or the magnetic field strength, with different evolutions of the SED at X-rays and $\gamma$-rays in the SSC model.
To investigate the evolution  at different flux levels, both
the  $\gamma$-ray and X-ray data are divided into four groups according to
the X-ray flux. A  hardening in the spectra toward  high fluxes is observed (see Figure~\ref{fig6} and Figure~\ref{fig7}).
The  results   based on two years of data are consistent with
the results obtained by the Whipple experiment \citep{krenn02}. A close tie between the variation of the flux and of the spectral index indicates  peak energy increases with the flux, which has also been  found in \citet[][]{aleksic10}. This supports the prediction of the SSC model for changes in the maximum momentum of emitting electrons \citep{MaK97}. Moreover, we study the correlation function between $\gamma$-rays and X-rays, finding that the $\gamma$-ray flux shows a quadratic increase with the X-ray flux (see Figure~\ref{fig8}). In the homogeneous SSC model, the synchrotron flux is proportional to the electron density, and the IC $\gamma$-ray flux is proportional to both the electron density and the synchrotron flux; therefore, the $\gamma$-ray flux is a quadratic function of the synchrotron flux.
Never before was there an indication distinguishing quadratic from linear correlations between $\gamma$-ray and X-ray fluxes according to observations of
flares, as reviewed by \citet[][]{wagner08}.
We also construct a  homogeneous one-zone
SSC model to simultaneously fit the $\gamma$-ray and X-ray emissions in four different flux levels (see Figure~\ref{fig9}) by
changing the electron parameters $l_e$ and/or $\gamma_{max}$ and/or
magnetic field strength. We find that the flux variation seems to be caused by the variation of the maximum energy and density of the electron injection
spectrum.

In conclusion, we have presented a long-term continuous
monitoring of Mrk 421 and a correlation between $\gamma$-rays observed by the
ARGO-YBJ experiment and satellite-borne X-ray data. The temporal and spectral
analysis strongly support the predictions of the SSC model.

\acknowledgments
 This work is supported in China by NSFC (No.10120130794),
the Chinese Ministry of Science and Technology, the
Chinese Academy of Sciences, the Key Laboratory of Particle
Astrophysics, and CAS, and in Italy by the Istituto Nazionale di Fisica
Nucleare (INFN).

We are grateful to Yupeng Chen for his help in estimating the X-ray spectrum using ASM/$RXTE$ data.
We also acknowledge the essential support of W. Y. Chen, G. Yang,
X. F. Yuan, C. Y. Zhao, R. Assiro, B. Biondo, S. Bricola, F. Budano,
A. Corvaglia, B. D'Aquino, R. Esposito, A. Innocente, A. Mangano,
E. Pastori, C. Pinto, E. Reali, F. Taurino, and A. Zerbini  in the
installation, debugging, and maintenance of the detector.
Furthermore, we thank the anonymous referee for the helpful comments and suggestions that improved the paper.

\clearpage

\begin{deluxetable}{cccccc}
\tablecolumns{3} \tablewidth{0pc} \tablecaption{Event Selections and the Number of Events}
 \tablehead{
\colhead{$N_{pad}$ range} & \colhead{$R$ (m)}   &
\colhead{TS  (ns$^2$)} & \colhead{Number of Events} }

\startdata
$[20,60]$ & No cut&  $<$50  & 8.71$\times$10$^{10}$ \\
$[60,100]$ & No cut& $<$30  & 1.83$\times$10$^{10}$\\
$[100,200]$ & $R<$70 & $<$20 & 6.13$\times$10$^{9}$ \\
$[200,500]$& $R<$70 & $<$20 & 3.42$\times$10$^{9}$ \\
$[500,1000]$& $R<$60 & $<$20  & 1.05$\times$10$^{9}$\\
$[>1000]$& $R<$30 & $<$20  & 3.71$\times$10$^{8}$ \\

\enddata
\end{deluxetable}

\begin{deluxetable}{cccccccc}
\tablecolumns{7} \tablewidth{0pc} \tablecaption{Best-Fit Parameters
in the SSC Model }
 \tablehead{
\colhead{Flux Level} & \colhead{$\gamma_{max}$}   &
\colhead{$l_{e}$} & \colhead{$B$ (G)} & \colhead{$R$ (cm)}    &
\colhead{$\delta$} & \colhead{$\alpha$}}

\startdata
1 &$7\times10^5$ &  $6\times10^{-6}$ &   0.08& $ 5\times10^{16}$&  16&  1.7 \\
2 &$7\times10^5$ &  $1\times10^{-5}$ &   0.15& $ 5\times10^{16}$&  15&  1.7 \\
3 &$1\times10^6$ &  $1\times10^{-5}$ &   0.15& $ 5\times10^{16}$&  15&  1.7 \\
4 &$2\times10^6$ &  $1.4\times10^{-5}$ &   0.15& $ 5\times10^{16}$&  15&  1.7 \\

\enddata
\end{deluxetable}

\clearpage

\clearpage

\end{document}